\documentclass{aastex}
\usepackage{spr-astr-addons}
\usepackage{url}

\usepackage{graphicx}
\usepackage{epsfig}
\RequirePackage{color}

\shorttitle{Bianchi type I cosmology}
 \shortauthors{Momeni et al.}

\begin{document}

\title{FRW and Bianchi type I cosmology of f-essence}

\author{M. Jamil \altaffilmark{1}}
\affil{Center for Advanced Mathematics and Physics, National
University of Sciences and Technology, Islamabad, Pakistan}

\author{D. Momeni\altaffilmark{2}}
\affil{Department of Physics, Faculty of sciences, Tarbiat Moa'llem
university, Tehran, Iran}

\author{N.S. Serikbayev\altaffilmark{3}}
\affil{Eurasian International Center for Theoretical Physics,
Eurasian National University, Astana, Kazakhstan}

\author{R. Myrzakulov\altaffilmark{3}}
\affil{Department of Physics, California State University, Fresno,
CA 93740 USA}

\altaffiltext{1}{mjamil@camp.nust.edu.pk}
\altaffiltext{2}{d.momeni@tmu.ac.ir; d.momeni@yahoo.com}
\altaffiltext{3}{rmyrzakulov@csufresno.edu; rmyrzakulov@gmail.com}

\begin{abstract}
F-essence is a generalization of the usual Dirac model with the
nonstandard kinetic term. In this paper,  we introduce a new model
of spinor cosmology containing both Ricci scalar and the non
minimally coupled spinor fields in its action. We have investigated
the cosmology with both isotropy and anisotropy, where the equations
of motion of FRW and Bianchi type-I spacetimes have been derived and
solved numerically. Finally the quantization of these models through
Wheeler-De Witt (WD) wave function has been discussed.
\end{abstract}


\section{Introduction}

Astrophysical data of supernovae of type Ia indicate that we live in
an accelerated expansion era of the Universe \citep{per99,rie98}.
There are two major theoretical explanations for this phenomena: the
first category is the fluid model  in which we keep the Einstein
gravity as a dominant theory and introduce a fluid in the right hand
side of the Einstein field equation and then investigate the
cosmological evolution of the model \citep{cop06} and \citep[paper
II]{ame00}. Another approach which has been investigated in recent
years is the geometrical one in which we search for generalized
models of the gravity which can deduce the accelerated expansion of
the Universe \citep[paper I]{eli04}. Some examples of later model is
$f(R)$ gravity \citep{noj03,cog08,aza08}, \citep[paper III]{eli04},
\citep[paper I]{jam11} and \citep[paper I]{mom09} in which the only
dynamical sector of action is a function of the Ricci scalar. Here
the non-linear terms of the curvature can be regarded as an
alternative for the accelerated expansion of the Universe. Another
class of models are the $f(R,G)$ models, where both curvature and
Gauss-Bonnet terms as the dynamical quantities \citep[paper
II]{eli04}, \citep{bam10}and \citep{fel10}. Another interesting
approach is the $f(T)$ gravity where $T$ is the torsion. Although it
has no curvature, the space-time manifold has time evolution, and
the dynamics has been caused by the Torsion only \citep{mol61,
hay67, wup10, pel63, yer10, che11,ben11,lib11, wei11}. Since its
equations of motions are lower order, working with it is easier,
also the matter comes from a non-minimally action and curvature.
This model can be explained as the trace of the specified
energy-momentum tensor, the common Ricci scalar, and the Lagrangian
$f(R,T)$ \citet{har11} and \citep[paper II]{mom09}. If we limit to
the scalar field theories, there are many options to have an
accelerated universe as quintessence \citep{zla99, ste99} and
\citep[paper I]{ame00},
quintom models \citep{fen06, zha06,q1,q2}, K-essense\citep{arm99}
and other combined models. There is no limitation for the usage of
the fermions in the matter sector of the theory as a matter source.
There are some works on spinor cosmology in the literature
\citep[last paper ]{arm99}, \citep[paper II]{vak05} and
\citep{rib05, cha07, boe08}. In this work, following an earlier work
on the generalization of the spinor cosmology \citep[paper
II]{jam11}, we have introduced a new model of spinor cosmology, in
which the action contains both $R$ and the non minimally coupled
spinor fields. We also have investigated both isotropic and
anisotropic cosmological models in this framework. We derive the
basic equations of motion and solve them for FRW and Bianchi type-I
models numerically. Finally both classical and quantum models in
this framework have been discussed and some analytical solutions for
the quantum cosmology have been studied.

\section{FRW metric in f-essence}

In this section we would like  to present the derivation of the
equations of motion for FRW metric in the f-essence.

Let us consider the following action of f-essence
\begin{equation} \label{}
S=\int d^{4}x\sqrt{-g}[R+2K( Y,   \psi, \bar{\psi})],
\end{equation}
where $R$ is the scalar curvature,    $Y$ is the kinetic term for
the fermionic field $\psi$ and  $K$ is some function of its
arguments. In the case of  the FRW  metric
\begin{equation} \label{}
ds^2=-dt^2+a^2(dx^2+dy^2+dz^2),
\end{equation}
 $R$ and  $Y$ have the form
\begin{equation} \label{}
 R=6\left(\frac{\ddot{a}}{a}+\frac{\dot{a}^2}{a^2}\right),
\end{equation}
\begin{equation} \label{}
Y=0.5i(\bar{\psi}\gamma^0\dot{\psi}-\dot{\bar{\psi}}\gamma^0\psi),
\end{equation}
respectively.
 Substituting these expressions  into (1) and integrating over the spatial dimensions,
  we are led to an effective Lagrangian in the mini-superspace $\{a, \psi, \bar{\psi}\}$
 \begin{equation}
L=-2(3a\dot{a}^2-a^3K).\end{equation}
 Variation of Lagrangian (5) with respect to $a$, yields the equation of motion
 of the scale factor
\begin{equation}
2a\ddot{a}+\dot{a}^2+a^2K=0.
\end{equation}

The variation of Lagrangian (5) with respect to $\bar{\psi}, \psi$
is the corresponding Euler-Lagrangian equations for the fermionic
fields
\begin{eqnarray}
&&K_Y\gamma^0\dot{\psi}+1.5\frac{\dot{a}}{a}K_Y\gamma^0\psi\nonumber\\&&+0.5\dot{K}_Y\gamma^0
\psi-iK_{\bar{\psi}}=0,
\\
&&K_Y\dot{\bar{\psi}}\gamma^0+1.5\frac{\dot{a}}{a}K_Y\bar{\psi}\gamma^0\nonumber\\&&+0.5
\dot{K}_Y\bar{\psi}\gamma^0+iK_{\psi}=0.
\end{eqnarray}
Another equivalence form is
\begin{eqnarray}
&&3H
K_{\dot{\psi}}+K_{\dot{\psi}\psi}\dot{\psi}+K_{\dot{\psi}\bar{\psi}}\dot{\bar{\psi}}\nonumber\\&&+
K_{\dot{\psi}\dot{\psi}}\ddot{\psi}+K_{\dot{\psi}\dot{\bar{\psi}}}\ddot{\bar{\psi}}+K_\psi=0,\\
&&3H
K_{\dot{\bar{\psi}}}+K_{\dot{\bar{\psi}}\psi}\dot{\psi}+K_{\dot{\bar{\psi}}\bar{\psi}}\dot{\bar{\psi}}\nonumber\\&&+
K_{\dot{\bar{\psi}}\dot{\psi}}\ddot{\psi}+K_{\dot{\bar{\psi}}\dot{\bar{\psi}}}\ddot{\bar{\psi}}+K_{\bar{\psi}}=0.
\end{eqnarray}
 Also the zero-energy condition is given by
\begin{equation}
L_{\dot{a}}\dot{a}+L_{\dot{\psi}}\dot{\psi}+L_{\dot{\bar{\psi}}}\dot{\bar{\psi}}-L=0,
\end{equation}
which yields the constraint
\begin{equation}
-3a^{-2}\dot{a}^2+YK_Y-K=0.
\end{equation}
Collecting all equations  and rewriting using the Hubble
 parameter $H=(\ln a)_t,$ we obtain a system of equations of f-essence
  (for the FRW metric case):
\begin{eqnarray}
&&3H^2-\rho=0,\\
&&2\dot{H}+3H^2+p=0,\\
&&3H
K_{\dot{\psi}}+K_{\dot{\psi}\psi}\dot{\psi}+K_{\dot{\psi}\bar{\psi}}\dot{\bar{\psi}}\nonumber\\&&+
K_{\dot{\psi}\dot{\psi}}\ddot{\psi}+K_{\dot{\psi}\dot{\bar{\psi}}}\ddot{\bar{\psi}}+K_\psi=0,\\
&&3H
K_{\dot{\bar{\psi}}}+K_{\dot{\bar{\psi}}\psi}\dot{\psi}+K_{\dot{\bar{\psi}}\bar{\psi}}\dot{\bar{\psi}}\nonumber\\&&+
K_{\dot{\bar{\psi}}\dot{\psi}}\ddot{\psi}+K_{\dot{\bar{\psi}}\dot{\bar{\psi}}}\ddot{\bar{\psi}}+K_{\bar{\psi}}=0,\\
&&\dot{\rho}+3H(\rho+p)=0.
\end{eqnarray}
Here
\begin{equation}
\rho=YK_{Y}-K,\quad p=K,
\end{equation}
are the energy density and the pressure of f-essence. It is clear
that these expressions for the energy density and the pressure
represent the components of the energy-momentum tensor of f-essence
as:
\begin{equation}
T_{00}=YK_{Y}-K,\quad  T_{11}=T_{22}=T_{33}=-K.
\end{equation}

\section{FRW model with $K( Y, \psi, \bar{\psi})=Y-V(\bar{\psi}\psi)$}

We introduce a useful model which is more applicable and more
suitable for exact solutions. This model is described by
\begin{equation}
K( Y, \psi, \bar{\psi})=Y-V(\bar{\psi}\psi).
\end{equation}
To obtain the field equations, we substitute this form in Eqs. (13-
17). Another simple method is re-deriving these equations using the
action directly, therefore in each of these equivalence methods we
have the following equations of motion for Dirac fields:
\begin{eqnarray}
3H(\frac{1}{2}i\bar{\psi}\gamma^0)+\frac{1}{2}i\gamma^0\dot{\bar{\psi}}-\frac{1}{2}i\dot
{\bar{\psi}}\gamma^0-V_\psi=0, \\
3H(-\frac{1}{2}i\gamma^0\psi)=V_{\bar{\psi}}.
\end{eqnarray}
FRW equations in this case are
\begin{eqnarray}
 &&2\frac{\ddot{a}}{a}+(\frac{\dot{a}}{a})^2+\frac{1}{2}i(\bar{\psi}\gamma^0\dot{\psi}-
 \dot{\bar{\psi}}\gamma^0\psi)\nonumber\\&&-V(\bar{\psi}\psi)=0 ,\\
&& 3H^2=V(\bar{\psi}\psi).
\end{eqnarray}
The general potential is $V(\bar{\psi}\psi)=2\bar{\psi}\psi$. For
this special case, we have the next set of EOMs:
\begin{eqnarray}
3H(\frac{1}{2}i\bar{\psi}\gamma^0)+\frac{1}{2}i\gamma^0\dot{\bar{\psi}}-\frac{1}{2}i
\dot{\bar{\psi}}\gamma^0-2\bar{\psi}=0,\\
-\frac{3}{2}iH\gamma^0\psi=2\psi,\\
 2\dot{H}+\frac{1}{2}i(\bar{\psi}\gamma^0\dot{\psi}-
 \dot{\bar{\psi}}\gamma^0\psi)=0.
\end{eqnarray}
Now we take the Dirac 2-spinor as $\bar{\psi}=(\psi_1,
\psi_2)^{\dag}\gamma^0$, the equation for spinor reads as
\begin{eqnarray}
\dot{\psi}+\frac{3}{2}H\psi+2i\gamma^ 0\psi=0.
\end{eqnarray}
Thus we must solve the next system of ODEs:
\begin{eqnarray}
\frac{ d\log\psi_a}{d t}=\frac{3}{2}H\psi_a\pm2,a=\{1,2\}=\{+,-\},
\end{eqnarray}
which posses the following solution
\begin{eqnarray}
\psi^{T}=(\psi_1(0)a(t)^{3/2}e^{2t},\psi_2(0)a(t)^{3/2}e^{-2t}).
\end{eqnarray}
Using this form of the 2-spinor we can obtain the scale factor from
the following equation
\begin{eqnarray}
\ddot{y}+i e^{3y}\{|\beta|^2e^{-4t}-|\alpha|^2e^{4t}\},
\end{eqnarray}
here $y\equiv
\log(a(t)),\{\alpha,\beta\}\equiv\{\psi_1(0),\psi_2(0)\}$. There is
no simple analytic solution for $y(t)$. But if we take
$\alpha=\beta=\frac{1}{\sqrt{i}}$, then we can solve it numerically.
FIG.1 shows the time evolution of $y(t)$ for some initial values.

\section{Bianchi type I cosmology of   f-essence}

The action of f-essence reads as
\begin{equation}
S=\int d^{4}x\sqrt{-g}[R+2K( Y,  \psi, \bar{\psi})],
\end{equation}
 where $K$ is some function of its arguments, $\psi=(\psi_1, \psi_2, \psi_3, \psi_4)^{T}$  is a
 fermionic function  and $\bar{\psi}=\psi^+\gamma^0$ is its adjoint function. Here
\begin{equation}
Y=0.5i[\bar{\psi}\Gamma^{\mu}D_{\mu}\psi-(D_{\mu}\bar{\psi})\Gamma^{\mu}\psi]
\end{equation}
is  the canonical kinetic term for the fermionic field and   $
D_{\mu}$ is covariant derivative
\begin{equation}
D_\mu\psi=\partial_\mu\psi+\Omega_\mu\psi, \quad
D_\mu\bar{\psi}=\partial_\mu\bar{\psi}-\bar{\psi}\Omega_\mu.
\end{equation}
Here $\Omega_\mu$ are spin connections, $\Gamma^{\mu}$ are the Dirac
matrices associated with the space-time metric satisfying the
Clifford algebra
\begin{equation}
\{\Gamma^{\mu},\Gamma^{\nu}\}=2g^{\mu\nu}.
\end{equation}
The $\Gamma^{\mu}$  are related to the flat Dirac matrices,
$\gamma^{a}$, through the tetrads $e^a_\mu$ as\begin{equation}
\Gamma^{\mu}=e^\mu_a\gamma^{a},\quad \Gamma_{\mu}=e^a_\mu\gamma_{a}.
\end{equation}
At the same time, the spin connections $\Omega_\mu$ satisfy the
relation
\begin{equation}
\Omega_{\mu}=0.25g_{nu\lambda}(\partial_\mu
e^{\lambda}_{a}+\Gamma^{\lambda}_{\sigma\mu}e^\sigma_a)\gamma^\nu\gamma^a.
\end{equation}
The tetrads can be easily obtained from their definition, that is
\begin{equation}
g_{\mu\nu}=e_\mu^ae^b_\nu\eta_{ab}.
\end{equation}
Let us now consider  the Bianchi type I  universe filled with
f-essence. These models for the special simple spinors have been
discussed previously \citep{sah04}. The metric is given by
\begin{equation}
ds^2=-N^2(t)dt^2+a^2(t)dx^2+b^2(t)dy^2+c^2(t)dz^2,
\end{equation}
where $a(t), b(t), c(t)$ are scale factors in the $x, y, z$
directions respectively and $N(t)$ is the lapse function. The
corresponding scalar curvature takes the form
\begin{eqnarray}
    R&=&\frac{2}{N^2}(\frac{\ddot{a}}{a}+\frac{\ddot{b}}{b}+\frac{\ddot{c}}{c}+
    \frac{\dot{a}\dot{b}}{ab}\nonumber\\&&+\frac{\dot{a}\dot{c}}{ac}+\frac{\dot{b}\dot{c}}{bc}-
    \frac{\dot{a}\dot{N}}{aN}-\frac{\dot{b}\dot{N}}{bN}-\frac{\dot{c}\dot{N}}{cN}),
    \end{eqnarray}
where a dot represents differentiation with respect to $t$. For the
metric (39) the tetrads take the form
    \begin{eqnarray}
    &&e^a_\mu=diag(N,a,b,c),\nonumber\\&&\quad
e_a^\mu=diag(1/N,1/a,1/b, 1/c).
\end{eqnarray}
These formulas yield
    \begin{eqnarray}
    &&\Omega_0=0,\quad \Omega_1=-\frac{\dot{a}}{2N}\gamma^0\gamma^1, \quad \Omega_2=
    -\frac{\dot{b}}{2N}\gamma^0\gamma^2,\nonumber\\&&\quad \Omega_3=-\frac{\dot{c}}{2N}\gamma^0\gamma^3,
\end{eqnarray}
 where $\gamma^0$ and $\gamma^i$ are the Dirac matrices in Minkowski spacetime and we have
 adopted the following representation
\begin{eqnarray}
\gamma^{0}  & = &
 \left(
\begin{array}{cc}
-i & 0 \\
0 & i
\end{array}     \right) ,
 \gamma^i= \left(
\begin{array}{cc}
0 &   \sigma^{i}\\
\sigma^{i} &   0
\end{array}     \right)
\end{eqnarray}
 Substituting (40) and (42) in (43) and integrating over the spatial dimensions, we are led
  to an effective Lagrangian in the mini-superspace $\{N,a,b,c, \psi, \bar{\psi}\}$
 \begin{equation}
L=-2[\frac{1}{N}(\dot{a}\dot{b}c+\dot{a}b\dot{c}+a\dot{b}\dot{c})-NabcK(
Y,  \psi, \bar{\psi})],\end{equation}
 where
 \begin{equation}
 L_f=2NabcK( Y,  \psi,\bar{\psi}),\quad  Y=\frac{1}{2N}(\bar{\psi}\gamma^{0}\dot{\psi}-\dot{\bar{\psi}}\gamma^{0}\psi).
  \end{equation}
 The preliminary set-up for writing the equations of motion is now complete.

Variation of Lagrangian (44) with respect to $N, a,b,c,\bar{\psi}$
and $\psi$ yields the equations of motion of the gravitational and
the fermions fields as:
    \begin{eqnarray}
&&\frac{\dot{a}\dot{b}}{ab}+\frac{\dot{a}\dot{c}}{ac}+\frac{\dot{b}\dot{c}}{bc}\nonumber\\&&-N^2(YK_Y-K)=0,\\
&&\frac{\ddot{b}}{b}+ \frac{\ddot{c}}{c}+ \frac{\dot{b}\dot{c}}{bc}\nonumber\\&&-\frac{\dot{N}}{N}(\frac{\dot{b}}{b}+\frac{\dot{c}}{c})+N^2K=0,\\
&&\frac{\ddot{a}}{a}+ \frac{\ddot{c}}{c}+ \frac{\dot{a}\dot{c}}{ac}\nonumber\\&&-\frac{\dot{N}}{N}(\frac{\dot{a}}{a}+\frac{\dot{c}}{c})+N^2K=0,\\
&&\frac{\ddot{b}}{b}+ \frac{\ddot{a}}{a}+ \frac{\dot{b}\dot{a}}{ba}\nonumber\\&&-\frac{\dot{N}}{N}(\frac{\dot{b}}{b}+\frac{\dot{a}}{a})+N^2K=0,\\
&&K_{Y}\dot{\psi}+0.5[(\ln(abc))_tK_{Y}+\dot{K}_{Y}]\psi\nonumber\\&&+N\gamma^0K_{\bar{\psi}}=0,\\
&&K_{Y}\dot{\bar{\psi}}+0.5[(\ln(abc))_tK_{Y}+\dot{K}_{Y}]\bar{\psi}\nonumber\\&&-NK_{\psi}\gamma^{0}=0,\\
&&  \dot{\rho}+3H(\rho+p)=0,
\end{eqnarray}
    where  energy density  and  pressure  take the form
\begin{equation}
\rho=N^2(YK_{Y}-K),\quad p=N^2K.
\end{equation}
The vacuum solutions of the above system i.e. $\psi=0$ is the
generalized Kasner solution in which space is homogeneous and has
Euclidean metric depending on time according to the Kasner metric
\citep{kas21}. This solution possessing a
Belinsky-Khalatnikov-Lifshitz (BKL) singularity \citep{bel69,lif60},
which is a model of dynamic evolution of the Universe near the
initial singularity $t=0$ and described by an anisotropic
homogeneous and chaotic solution to the Einstein's field equations
of gravitation. The Mixmaster universe exhibits similar properties
as the Kasner solution.

 Some properties of g-essence were studied in \citep{kul11,raz11}.
Model (1) admits two important reductions: \emph{k-essence} and
\emph{f-essence}.

\section{Bianchi type I cosmologies with $K( Y, \psi, \bar{\psi})=Y-V(\bar{\psi}\psi)$}

In this section we examine the Bianchi type I cosmology for $K( Y,
\psi, \bar{\psi})=Y-V(\bar{\psi}\psi)$. In this case we
have
\begin{eqnarray}
&&\frac{\dot{a}\dot{b}}{ab}+\frac{\dot{a}\dot{c}}{ac}+\frac{\dot{b}\dot{c}}{bc}\nonumber\\&&-N^2V(\bar{\psi}\psi)=0,\\
&&      \frac{\ddot{b}}{b}+ \frac{\ddot{c}}{c}+
\frac{\dot{b}\dot{c}}{bc}\nonumber\\&&-\frac{\dot{N}}{N}(\frac{\dot{b}}{b}
        +\frac{\dot{c}}{c})+N^2K=0,\\
       && \frac{\ddot{a}}{a}+ \frac{\ddot{c}}{c}+ \frac{\dot{a}\dot{c}}{ac}-
        \frac{\dot{N}}{N}(\frac{\dot{a}}{a}+\frac{\dot{c}}{c})\nonumber\\&&+N^2K=0,\\
     &&   \frac{\ddot{b}}{b}+ \frac{\ddot{a}}{a}+ \frac{\dot{b}\dot{a}}{ba}
        -\frac{\dot{N}}{N}(\frac{\dot{b}}{b}+\frac{\dot{a}}{a})\nonumber\\&&+N^2K=0,\\
         &&       \dot{\psi}+0.5(\ln(abc))_t\psi\nonumber\\&&-N\gamma^0V_{\bar{\psi}}=0,\\
&&\dot{\bar{\psi}}+0.5(\ln(abc))_t\bar{\psi}\nonumber\\&&+NV_{\psi}\gamma^{0}=0,\\
 &&   \dot{\rho}+3H(\rho+p)=0.
    \end{eqnarray}
We take $V(\bar{\psi}\psi)=2 \bar{\psi}\psi$. Thus we obtain
\begin{eqnarray}
&&\frac{\dot{a}\dot{b}}{ab}+\frac{\dot{a}\dot{c}}{ac}+\frac{\dot{b}\dot{c}}{bc}
\nonumber\\&&-2N^2 \bar{\psi}\psi=0,\\
&&\frac{\ddot{b}}{b}+ \frac{\ddot{c}}{c}+
\frac{\dot{b}\dot{c}}{bc}-\frac{\dot{N}}{N}(\frac{\dot{b}}{b}
+\frac{\dot{c}}{c})\nonumber\\&&+N^2(\frac{1}{2N}(\bar{\psi}\gamma^{0}\dot{\psi}-\dot{\bar{\psi}}
\gamma^{0}\psi)-2 \bar{\psi}\psi)=0,\\
&&\frac{\ddot{a}}{a}+ \frac{\ddot{c}}{c}+ \frac{\dot{a}\dot{c}}{ac}-
\frac{\dot{N}}{N}(\frac{\dot{a}}{a}+\frac{\dot{c}}{c})\nonumber\\&&+N^2(\frac{1}{2N}(\bar{\psi}\gamma^{0}
\dot{\psi}-\dot{\bar{\psi}}\gamma^{0}\psi)-2 \bar{\psi}\psi)=0,\\
&&\frac{\ddot{b}}{b}+ \frac{\ddot{a}}{a}+ \frac{\dot{b}\dot{a}}{ba}
-\frac{\dot{N}}{N}(\frac{\dot{b}}{b}+\frac{\dot{a}}{a})\nonumber\\&&+N^2(\frac{1}{2N}
(\bar{\psi}\gamma^{0}\dot{\psi}-\dot{\bar{\psi}}\gamma^{0}\psi)-2 \bar{\psi}\psi)=0,\\
&&\dot{\psi}+0.5(\ln(abc))_t\psi-2N\gamma^0\psi=0,\\
&&\dot{\bar{\psi}}+0.5(\ln(abc))_t\bar{\psi}+2N\bar{\psi}\gamma^{0}=0,\\
&&\dot{\rho}+3H(\rho+p)=0.
    \end{eqnarray}
We solved these equations numerically for a set of initial
conditions imposed on the set of the functions
$\{a(t),b(t),c(t),|\psi(t)|^2\}$. We set $N(t)=1$ without loss of
generality, since the metric is a projectable metric, i.e. we can
define a new time coordinate $t'=\int N(t) dt$. The gauge $N(t) = 1$
chosen in classical cosmological models, and called the cosmic time
gauge. Another gauge fixing leads to $N_i = 0$, here $N_i$ is the
shift vector.

The numerical solutions is shown in the FIG.2. As we observe, the
functions $\{a(t),b(t),c(t)\}$ are monotonically increasing
functions of time, but the density function $|\psi(t)|^2$ is a
decreasing function of $t$.

\section{Quantization of f-essence }

For quantization of the model as described in (32) we adopt the
method proposed by Misner \citep{mis69}. The first step is writing
the general Hamiltonian suitable for describing the quantum
evolution of the system. For a typical model (32), with an unknown
form of the function $K$, it is not possible to write such
Hamiltonian. But if we restrict to the case $K( Y, \psi,
\bar{\psi})=Y-V(\bar{\psi}\psi)$, the problem at hand becomes
tractable. This is a special case of the form discussed previously
in \citep[paper I]{vak05}. First we introduce a set of the metric
functions
\begin{eqnarray}
 \chi=\frac{1}{2\sqrt{3}}\log(\frac{a}{b}),\\
 \rho=\log(\sqrt{c\sqrt{ab}}),\\
 \sigma=\frac{1}{2}\log(\frac{\sqrt{ab}}{c}).
  \end{eqnarray}
It is easy to show that the Hamiltonian for model $K( Y, \psi,
\bar{\psi})=Y-V(\bar{\psi}\psi)$ is
\begin{eqnarray}
H=\frac{e^{-3\rho}}{12}(p_\rho^2-p_\sigma^2-p_\chi^2)
+e^{3\rho}[V(\bar{\psi}\psi)-\Upsilon]=0,
  \end{eqnarray}
where $\Upsilon$ is the Lagrange multiplier of the system. The set
of the corresponding conjugate momentums of the new set of
configurational coordinates $\{\rho,\sigma,\chi\}$ is
$\{p_\rho,p_\sigma,p_\chi\}$ which satisfy the commutation brackets.
With a specified form of the interaction $V(\bar{\psi}\psi)$, we can
obtain the classical solutions described in \citep[paper I]{vak05}.
Now from (71), we get the wave function directly from the WD
equation \citep{dew67}, and with the usual replacements
$p_i\rightarrow -i \frac{\partial}{\partial x_i}$, the Wheeler-De
Witt equation is
\begin{eqnarray}
\Big[\frac{1}{12}(-\partial_\rho^2+\partial_\sigma^2+\partial_\chi^2)
+e^{6\rho}(-e^{-3\rho}-\Upsilon)\Big]\Psi(\rho,\sigma,\chi)=0.
  \end{eqnarray}
We write the wave function as $\Psi(\rho,\sigma,\chi)=e^{i(k_\sigma
\sigma+k_\chi \chi)}\Gamma(\rho)$ where
\begin{eqnarray}
-\partial_\rho^2\Gamma(\rho)+[-k_\sigma ^2-k_\chi ^2+12
e^{6\rho}(-e^{-3\rho}-\Upsilon)]\Gamma(\rho)=0.
  \end{eqnarray}
The general solution for (73) is
\begin{eqnarray}
\Gamma(\rho)={{\rm e}^{-3/2\,\rho}} \Big[ {\it C_1}\, {{\rm \bf
M}\left({\frac {-1/3\,i\sqrt {3}}{\sqrt
{\Upsilon}}},\,1/3\,ik,\,4/3\,i\sqrt {3}\sqrt {\Upsilon}{{\rm
e}^{3\,\rho}}\right)} \\ \nonumber+{\it C_2}\, {{\rm \bf
W}\left({\frac {-1/3\,i\sqrt {3}}{\sqrt
{\Upsilon}}},\,1/3\,ik,\,4/3\,i\sqrt {3}\sqrt {\Upsilon}{{\rm
e}^{3\,\rho}}\right)}\Big].
\end{eqnarray}
Here $M$ and $W$ are $WhittakerM$ and $WhittakerW$ functions
respectively while $k=\sqrt{k_\sigma ^2+k_\chi ^2}$. Thus the total
wave function is
\begin{eqnarray}
\Psi(\rho,\sigma,\chi)&=&\sum_{k_\sigma,k_\chi}e^{i(k_\sigma
\sigma+k_\chi \chi)}{{\rm e}^{-3/2\,\rho}} \\
\nonumber&&\times\Big[ {\it a_{k}}\, {{\rm \bf M}\Big({\frac
{-1/3\,i\sqrt {3}}{\sqrt {\Upsilon}}},\,1/3\,ik,\,4/3\,i\sqrt
{3}\sqrt {\Upsilon}{{\rm e}^{3\,\rho}}\Big)}\\ \nonumber&& +{\it
b_k}\, {{\rm \bf W}\Big({\frac {-1/3\,i\sqrt {3}}{\sqrt
{\Upsilon}}},\,1/3\,ik,\,4/3\,i\sqrt {3}\sqrt {\Upsilon}{{\rm
e}^{3\,\rho}}\Big)}\Big].
\end{eqnarray}
For normalization we set $b_k=0$ and truncate the series for
convergence. Thus the solution is written in final form as
\begin{eqnarray}
\Psi(\rho,\sigma,\chi)&=&\sum_{k_\sigma,k_\chi}e^{i(k_\sigma
\sigma+k_\chi \chi)}{{\rm e}^{-3/2\,\rho}}  {\it a_{k}}\, \\
\nonumber&&\times{{\rm \bf M}\Big({\frac {-1/3\,i\sqrt {3}}{\sqrt
{\Upsilon}}},\,1/3\,ik,\,4/3\,i\sqrt {3}\sqrt {\Upsilon}{{\rm
e}^{3\,\rho}}\Big)}.
\end{eqnarray}
The Fourier-Whittaker coefficients $a_k$ can be obtained from the
initial wave function $\Psi(0,\sigma,\chi)$. For Gaussian wave
packet we can obtain the following result for the Fourier amplitude
as follow:
\begin{equation}
a_k = \frac{e^{-\frac{k^2}{2}}}{2 \pi }. \label{coef}
\end{equation}
In Eq.~(\ref{coef}) we use the approximation $\Upsilon \ll 1$.

\section{Conclusion}

In conclusion, we derived the equations of motion of f-essence for
FRW and Bianchi type I metrics. It is shown that if the Lagrangian
of fermionic fields $K$ has
 the usual Dirac form than the corresponding
 results coincide with the standard Einstein-Dirac theory. We have
 investigated both classical and quantum aspects of this model.

\section*{Acknowledgments}
R. Myrzakulov would like to  thank   D. Singleton  and Department of
Physics, California State University Fresno
  for their hospitality during his one year visit (October, 2010 -- October, 2011).

\section*{Appendix}

As a double check, one can obtain the above field equations from the
Einstein and Dirac equations given by:
 \begin{eqnarray}
    R_{\mu\nu}-\frac{1}{2}Rg_{\mu\nu}&=&T_{\mu\nu},\\
        \Gamma^{\mu}D_{\mu}\psi+K_{\bar{\psi}}&=&0,\\
        D_{\mu}\bar{\psi}\Gamma^{\mu}+K_{\psi}&=&0,\\
    \dot{\rho}+3H(\rho+p)&=&0.
    \end{eqnarray}
 For a homogeneous fermionic field $\psi(t)$, equations (A2) and (A3)
  are equivalent to (A1) and (A4)
 respectively. On the other hand, the non-vanishing components of the
  Einstein tensor for the metric (39)
 are:
 \begin{eqnarray}
G_{00}&=&\frac{\dot{a}\dot{b}}{ab}+\frac{\dot{a}\dot{c}}{ac}+\frac{\dot{b}\dot{c}}{bc},\\
G_{11}&=&-\frac{a^2}{N^2}\Big[\frac{\ddot{b}}{b}+   \frac{\ddot{c}}{c}+
\frac{\dot{b}\dot{c}}{bc}-\frac{\dot{N}}{N}(\frac{\dot{b}}{b}+\frac{\dot{c}}{c})\Big],\\
        G_{22}&=&-\frac{b^2}{N^2}\Big[\frac{\ddot{a}}{a}+   \frac{\ddot{c}}{c}+ \frac{\dot{a}\dot{c}}{ac}-\frac{\dot{N}}{N}\Big(\frac{\dot{a}}{a}+\frac{\dot{c}}{c}\Big)\Big],\\
        G_{33}&=&-\frac{c^2}{N^2}\Big[\frac{\ddot{b}}{b}+   \frac{\ddot{a}}{a}+ \frac{\dot{b}\dot{a}}{ba}-\frac{\dot{N}}{N}\Big(\frac{\dot{b}}{b}+\frac{\dot{a}}{a}\Big)\Big].
    \end{eqnarray}
The components of the energy-momentum tensor for the fermionic field
as the matter source can be obtained from the standard definition
as:
\begin{equation}
  T_{\mu\nu}=2\frac{\partial L_f}{\partial g^{\mu\nu}}-g_{\mu\nu}L_f,
  \end{equation}
yielding
\begin{eqnarray}
 && T_{00}=-2N^2(YK_Y-K),\quad T_{11}=-2a^2K, \quad T_{22}=-2b^2K,
 \nonumber\\&& \quad T_{33}=-2c^2K, \quad T_{ij}=T_{0i}=0.
  \end{eqnarray}
Substituting these results into Einstein equations (A1), yields the
same equations as (46) -(52). In the case of the FRW metric (3), the
equations corresponding to the action (1)
    can be obtained as:
    \begin{eqnarray}
    3H^2-\rho&=&0,\\
        2\dot{H}+3H^2+p&=&0,\\
                K_{Y}\dot{\psi}+0.5(3HK_{Y}+\dot{K}_{Y})\psi-i\gamma^0K_{\bar{\psi}}&=&0,\\
K_{Y}\dot{\bar{\psi}}+0.5(3HK_{Y}+\dot{K}_{Y})\bar{\psi}+iK_{\psi}\gamma^{0}&=&0,\\
    \dot{\rho}+3H(\rho+p)&=&0,
    \end{eqnarray}
where  the kinetic terms, the energy density  and  the pressure take
the forms
\begin{equation}
 Y=0.5i(\bar{\psi}\gamma^{0}\dot{\psi}-\dot{\bar{\psi}}\gamma^{0}\psi),
  \end{equation}

  \begin{equation}
  \rho=K_{Y}Y-K,\quad
p=K.
  \end{equation}

\begin{figure}
 \includegraphics[scale=0.3]{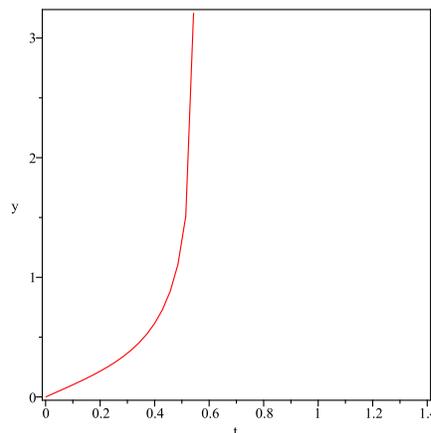}
  \caption{ Numerical solution for $y(t)=\log (a(t))$ .}
 \label{1.eps}
\end{figure}
\begin{figure}
 \includegraphics[scale=0.4]{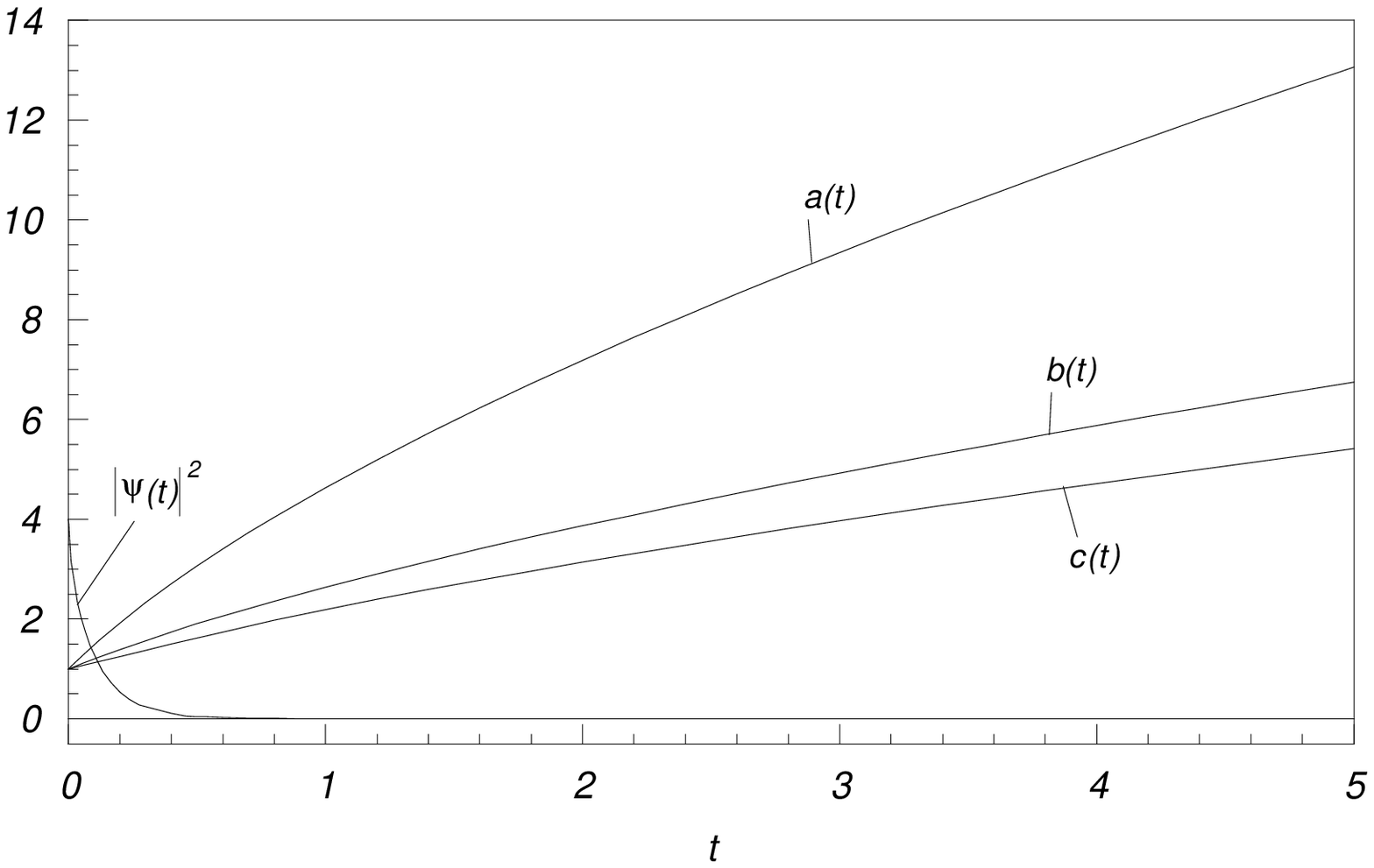}
  \caption{ Numerical solution for $\{a(t),b(t),c(t),|\psi(t)|^2\}$ .}
 \label{2.eps}
\end{figure}

\end{document}